\documentclass[12pt]{article}%
\usepackage{amsmath}
\usepackage{graphicx}
\usepackage{amsfonts}
\usepackage{amssymb}%
\providecommand{\U}[1]{\protect\rule{.1in}{.1in}}
\begin{document}

\title{The Heart of Fidelity}
\author{Gerd Bergmann\\Department of Physics \& Astronomy\\University of Southern California\\Los Angeles, California 90089-0484\\e-mail: bergmann@usc.edu}
\date{\today }
\maketitle

\begin{abstract}
The multi-electron wave function of an interacting electron system depends on
the size of the system, i.e. the number of electrons. Here the question
investigated is how the wave function changes for a symmetric Friedel-Anderson
impurity when the volume is doubled. It turns out that for sufficiently large
volume (when the level spacing is smaller than the resonance width) the change
in the wave function can be expressed in terms of a universal single-electron
state $\left\vert q\right\rangle $ centered at the Fermi level. This electron
state is independent of the number of electrons and independent of the
parameters of the Friedel-Anderson impurity. It is even the same universal
state for a Kondo impurity and a symmetric Friedel impurity independent of any
parameter. The only requirement is that the impurity has a resonance exactly
at the Fermi level and that the level spacing is smaller than the resonance
width. This result clarifies recent fidelity calculations.


\end{abstract}

In the late 1960s Anderson \cite{A53} showed that the potential of a weak
impurity in a metal host changes the total $n$-electron wave function of the
conduction electrons dramatically. Actually with increasing number $N_{c}$ of
electron states (which is achieved by increasing the volume) the scalar
product between the wave functions of the $n$-electron host without and with
the impurity approaches zero. This phenomenon is generally called the Anderson
orthogonality catastrophe (AOC). In recent years this phenomenon has been
somewhat generalized and decorated with the romantic name fidelity. The
generalization is that one applies the AOC to an arbitrary system which
depends on one or several parameters $\lambda$. If the system consists of
electrons then it is described by its Hamiltonian. The Hamiltonian may contain
a term which is proportional to a parameter $\lambda$. Suppose that one can
calculate the ground state of the system for $\lambda=0$ and for finite
$\lambda$. Then the scalar product of the two wave functions is defined as the
fidelity $F_{N_{c}}\left(  0,\lambda\right)  $ of the system. Here $N_{c}$ is
the number of conduction electrons states, which is proportional to the
volume. The fidelity depends on the size of the system and of particular
interest is the limit for $N_{c}$ increasing towards infinity. If $F_{N_{c}%
}\left(  0,\lambda\right)  $ approaches zero in this limit (the thermodynamic
limit), then one faces an AOC.

Our group studied recently the fidelity of the Friedel-Anderson impurity. This
is an electron system with a d-atom as impurity. The energy of the d-electron
lies at $E_{d}$ below the Fermi level. If one removes a d-electron, i.e.
creates a d-hole, then the conduction electrons can hop into the empty d-state
with a hopping matrix element $V_{sd}$. The d-hole possesses a finite life
time $\tau_{d}$ before before it is refilled. Due to Heisenberg's uncertainty
principle this life time broadens the energy level of the d-electron and
transforms it into a d-resonance with a resonance width which is of the order
of $\hbar/\tau_{d}$. In general the properties of a dissolved d-atom are more
complicated because the different d-electrons repel each other due to the
Coulomb interaction. In the theoretical investigation of such an impurity one
studies (most of the time) a simplified model of a d-impurity with only two
d-states, a spin-up and a spin-down d-state. Such an impurity was first
studied by Friedel \cite{F28} and Anderson \cite{A31} and I call it the
Friedel-Anderson (FA) impurity. The strength of the Coulomb interaction $U$
represents a parameter $\lambda$ as introduced in the fidelity. For $U=0$ the
impurity properties are much simpler. The impurity has a Friedel resonance at
the d-energy $E_{d}$ in each spin sub-band and is called a Friedel impurity.

While the wave function of the FA impurity is quite complex the density of
states is simpler and qualitatively sketched in Fig.1 together with the
density of states of a Friedel impurity. For both impurities the symmetric
case is shown. In the Friedel impurity the resonance is positioned at the
Fermi level ($E_{d}=0$). In the FA impurity the d-state energy is positioned
at $E_{d}=-U/2$ so that $E_{d}$ and $\left(  E_{d}+U\right)  $ lie
symmetrically to the Fermi level. As a consequence there is an electron-hole
symmetry. One obtains two d-resonances at roughly the energies $E_{d}=-U/2$
and $E_{d}+U=+U/2$ for spin up and down. These are known as the Hubbard
resonances, and their width is twice the width of a Friedel resonance with the
same s-d-hopping \cite{S88}, \cite{L57}, \cite{B181}. In addition one obtains
a narrow resonance at the Fermi level which is generally called the Kondo
resonance (see for example ref. \cite{H20}).%

\begin{align*}
&
{\includegraphics[
height=2.5131in,
width=4.8418in
]%
{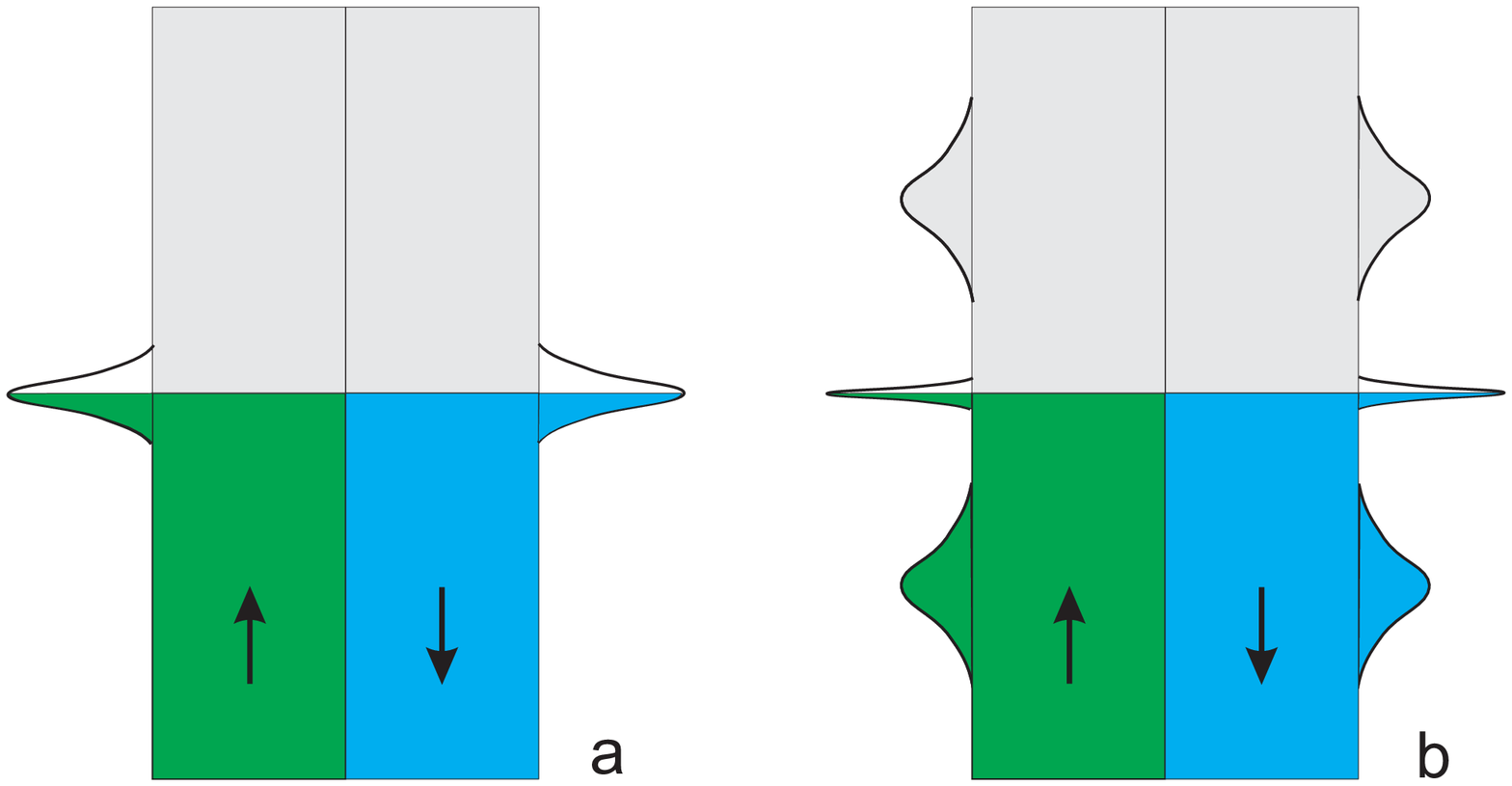}%
}%
\\
&
\begin{tabular}
[c]{l}%
Fig.1: a) The electron band of a symmetric Friedel impurity with\\
the d-resonance at the Fermi level.\\
b) The effective band (density of states) after turning on the Coulomb\\
repulsion $U$ and setting $E_{d}=-U/2$. The Friedel resonances are\\
transformed into broad Hubbard resonances, roughly positioned at\\
$E_{d}=-U/2$ and $E_{d}+U=+U/2$, far away from the Fermi level\\
and an extremely narrow Kondo resonance at the Fermi level.
\end{tabular}
\end{align*}%
\[
\]

Two studies of the fidelity of the FA impurity have been published recently,
one by Weichselbaum et al. \cite{W50} and one by our group \cite{B185}. In our
investigation we calculated the fidelity between a symmetric Friedel impurity
($E_{d}=0$, $U=0$) and a symmetric FA impurity with finite Coulomb repulsion
$U$ and $E_{d}=-U/2$. When level spacing $\delta E$ is smaller than the width
of the Kondo resonance in Fig.1b then the fidelity did not change any more
with increasing $N$. This raised a puzzled question by one referee of our
paper: 'Does this mean that the simple Friedel resonance and the complex FA
resonance have the same wave function (close to the Fermi level)'. In a way
this question stimulated the present investigation.

In the fidelity calculations one has on one hand to increase the number of
electron states dramatically. On the other hand one needs to keep the number
of states relatively small because otherwise the numerics requires an
unacceptable computer time. These opposing requirements are optimally
fulfilled by an ingenious trick applied by Wilson. One considers a system with
$2^{N/2}$ electron states. The conduction band is half filled and symmetric to
the Fermi level. For simplicity one assumes a constant density of states and
divides all energies by the Fermi energy. Then the conduction band extends
from $\left(  -1:+1\right)  $ as shown in Fig.2. In the next step one divides
the lower (and upper) half of the band geometrically into cells with
decreasing width so that one obtains an energy frame. In Fig.2 this energy
frame has the values $-1,-1/2,-1/4,$ $-1/8,-1/16,-1/32,$ $0,1/32,1/16,$
$1/8,1/4,1/2,1$ and defines $N$ energy cells $\mathfrak{C}_{\nu}$ (in Fig.2 we
have $N=12$). The number $N$ is always even because there are as many negative
as positive Wilson states. Each sub-division adds two states so that the
number of sub-divisions or iterations is equal to $N/2$ .

The four energy cells close to the Fermi level (in Fig.2 $\mathfrak{C}%
_{5},\mathfrak{C}_{6},\mathfrak{C}_{7},\mathfrak{C}_{8}$) may be considered to
possess just one electron state each. The number of states per cell doubles
with each step away from the Fermi level. (In Fig.2 $\mathfrak{C}_{1}$ and
$\mathfrak{C}_{12}$ have 16 electron states and the whole band has 64). Wilson
reshuffled these electron states in each cell in such a way that one of them,
the state $c_{\nu}^{\dagger}$ in the cell $\mathfrak{C}_{\nu}$, carried the
full interaction with the d-impurity. The remaining states in the cell have
zero interaction and are neglected. The width of the four smallest energy
cells in the vicinity of the Fermi level determines the effective energy
spacing $\delta E=2^{-\left(  N/2-1\right)  }$ (in units of the Fermi energy).
The effective number of electron states $N_{eff}$ is determined by the energy
spacing at the Fermi level and is given by $N_{eff}=2/\delta E$=2$^{N/2}$ (for
details see the discussion in ref.\cite{B185}). According to the construction
the number $N$ of Wilson states in FAIR is always even.

The Wilson states have two great advantages: (i) One can represent a band of
$2^{N/2}$ electrons by just $N$ Wilson states and (ii) One can double the
number of electrons by subdividing the cell directly below and the cell
directly above the Fermi level into two equal halves. So the Wilson spectrum
achieves the trick at doubling the effective number of states by adding just
two states.
\[%
\begin{array}
[c]{cc}%
{\includegraphics[
height=3.2619in,
width=1.5268in
]%
{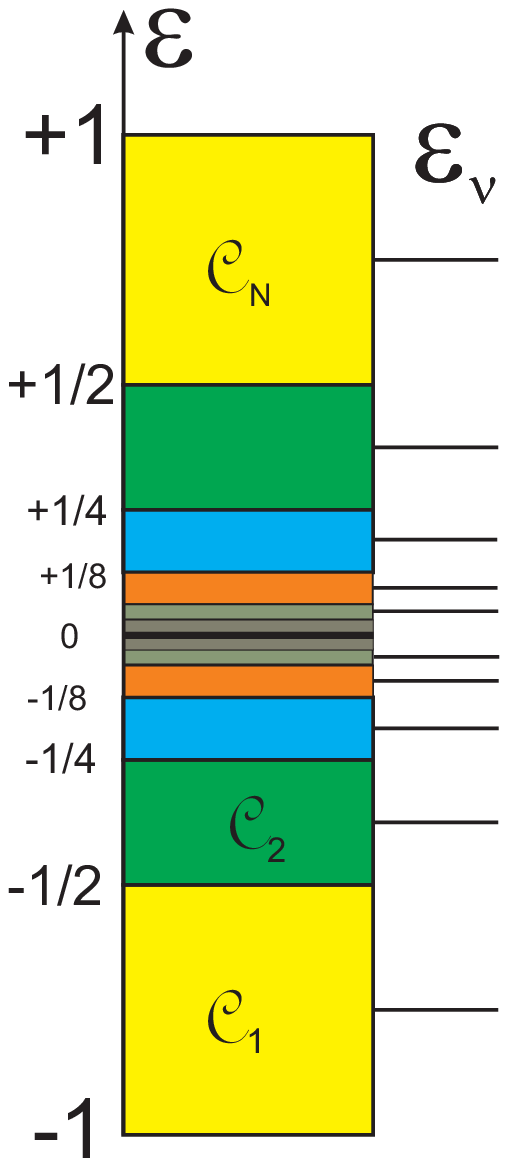}%
}%
&
\begin{tabular}
[c]{l}%
Fig.2: The Wilson description of a conduction band.\\
The energies are normalized with the Fermi energy.\\
The density of states is constant. The lower and upper\\
half are subdivided into energy cells whose size is\\
shrinks by $\Lambda=2$ towards the Fermi level (at zero\\
energy).
\end{tabular}
\end{array}
\]%
\[
\]

In this investigation I ask the question: Suppose that for a half-filled
$\ $band with $\left(  N-2\right)  $ Wilson states we calculate the ground
state $\Psi_{N-2}$ of a given impurity and then repeat the calculation for a
system with $N$ Wilson states for the same physical parameters, again half
filled. The occupation of each spin band is either $\left(  N-2\right)  /2$ or
$N/2$ which I define as $n=N/2.$ The ground state with $N$ Wilson states has
one additional spin-up electron and one additional spin-down electron. The
ground state $\Psi_{N-2}$ can be easily expressed in the new basis. If we
denote for a moment the two energy cells next to the Fermi level in the
$\left(  N-2\right)  $-basis as $\mathfrak{C}_{-}$ and $\mathfrak{C}_{+}$ then
these cells are each split into half in the $N$-basis. Consequently the
amplitude in each of the new states is just $1/\sqrt{2}$ of the amplitude in
the original basis state. So the state $\Psi_{N-2}$ is exactly transferred
into the $N$-state Wilson basis.

The question in this paper is the following: What is the relation between
$\Psi_{N-2}$ and $\Psi_{N}$ in the range of $N$ where the fidelity is
constant? I arrive at the following result: For sufficiently large $N$ there
is a single electron state $q^{\dagger}$ such that $\Psi_{N}\simeq
q_{\uparrow}^{\dagger}q_{\downarrow}^{\dagger}\Psi_{N-2}$ with remarkable
accuracy. This conclusion will be derived in detail below. (In this paper I
denote single-electron states such as $\left\vert q\right\rangle $ by their
creation operator $q^{\dagger}$).

The Kondo and the FA-impurity problem have been exactly solved with the Bethe
ansatz \cite{W12}, \cite{S29}. I am told that it is very hard to extract the
wave function from the Bethe-ansatz. The presently most frequently used
numerical method for the investigation of the FA impurity is the numerical
renormalization group (NRG) theory, which was developed by Wilson \cite{W18}
35 years ago and first applied to the FA impurity by Krishna-murthy et al.
\cite{K58} in 1980. It derives the ground state through a large number of
renormalization steps (of the order of 50 to 100 steps). In each step the
number of Slater states is increased by a factor of 16. (A Slater state is
defined as the product of $n$ single electron states). \ This yields a huge
number of Slater states for the ground state and only a small number of the
order of a few thousand Slater states are finally included in the calculation performed.

For the actual calculation I use the FAIR ground state for the different
impurities \cite{B91}, \cite{B151}, \cite{B153}. The FAIR technique has been
developed during the past few years by the author. FAIR stands for Friedel
artificially inserted resonance. The FAIR ground state represents a very good
approximation for the Friedel-Anderson and the Kondo impurities. It has
reproduced a number of numerical results with good accuracy and produced a
number of new results, such as the polarization of the Kondo cloud
\cite{B177}, oscillations in the Kondo cloud \cite{B182}, Friedel oscillations
of the FA impurity \cite{B186}, and, in the magnetic pseudo-ground state, the
magnetic moment \cite{B152} which roughly corresponds to the mean field result
with half the Coulomb energy $U$ because of the reduced density of states in
the d-resonances \cite{B181}. A review of the FAIR theory is given in ref.
\cite{B187}.

The singlet ground state of the FA impurity, which consists of eight Slater
states, was used to calculate the fidelity between the symmetric Friedel and
the symmetric FA impurity \cite{B185}. Its wave function is essentially the
superposition of two magnetic states $\Psi_{MS}$ with opposite magnetic
moment. The magnetic state (with net spin up) has the form
\begin{equation}
\Psi_{MS\uparrow}=\left[  Aa_{0\uparrow}^{\dag}b_{0\downarrow}^{\dag
}+Ba_{0\uparrow}^{\dag}d_{\downarrow}^{\dag}+Cd_{\uparrow}^{\dag
}b_{0\downarrow}^{\dag}+Dd_{\uparrow}^{\dag}d_{\downarrow}^{\dag}\right]
\left\vert \mathbf{0}_{a\uparrow}\mathbf{0}_{b\downarrow}\right\rangle
\label{Psi_MS}%
\end{equation}
where $a_{0\uparrow}^{\dagger}$ and $b_{0\downarrow}^{\dagger}$ are two
artificial resonance states or \textbf{fair} states in the spin-up and down
bands. It is defined and investigated in ref. \cite{B152} and \cite{B181}
within FAIR formalism. Its magnetic moment is $\left\vert B^{2}-C^{2}%
\right\vert $ in units of $\mu_{B}$.

The composition, for example, of $a_{0\uparrow}^{\dagger}$ in terms of the $N$
Wilson states $c_{\nu\uparrow}^{\dagger}$ is
\[
a_{0\uparrow}^{\dagger}=%
{\displaystyle\sum_{\nu=1}^{N}}
\alpha_{0}^{\nu}c_{\nu\uparrow}^{\dagger}%
\]
The \textbf{fair }states assume the effective interaction with the impurity.
Since the electron system has two spin sub-bands one needs two \textbf{fair}
states, the state $a_{0\uparrow}^{\dagger}$ for the spin-up sub-band and the
state $b_{0\downarrow}^{\dagger}$ for the spin-down sub-band. The remaining
states $a_{i\uparrow}^{\dagger}$ in the spin-up sub-band are made ortho-normal
to each other and to $a_{0\uparrow}^{\dagger}.$ In addition their free
electron Hamiltonian matrix $\left\langle a_{i\uparrow}^{\dagger}%
\Omega\left\vert H^{0}\right\vert a_{j\uparrow}^{\dagger}\Omega\right\rangle $
is sub-diagonalized (excluding the row and column with $a_{0\uparrow}%
^{\dagger}$ matrix elements). As a result the state $a_{0\uparrow}^{\dagger}$
becomes an artificial Friedel resonance. The \textbf{fair} state
$a_{0\uparrow}^{\dagger}$ (and $b_{0\downarrow}^{\dagger}$) determines
uniquely the remaining band states $a_{i\uparrow}^{\dagger}$ (and
$b_{i\downarrow}^{\dagger}$) for $i\geq1$ which form a new conduction band
with one less electron (for each spin). The half-occupied bands are
represented by
\[
\left\vert \mathbf{0}_{a\uparrow}\mathbf{0}_{b\downarrow}\right\rangle
=\prod_{i=1}^{n-1}a_{i\uparrow}^{\dag}\prod_{i=1}^{n-1}b_{i\downarrow}^{\dag
}\Phi_{0}%
\]
where $n=N/2$.

The FA ground state is the normalized superposition of the magnetic states
with net spin up and net spin down.%
\begin{equation}
\Psi_{SS}=\Psi_{MS\uparrow}+\Psi_{MS\downarrow}%
\end{equation}
The state $\Psi_{MS\downarrow}$ is obtained by reversing all spins in equ.
(\ref{Psi_MS}) (the spins are ordered in the same fashion as in equ.
(\ref{Psi_MS})). The main numerical task is to find the optimal \textbf{fair}
states $a_{0}^{\dagger}$ and $b_{0}^{\dagger}$ (which occur now in both spin
directions). When this is achieved by variation the full wave function can be
easily constructed.

The FAIR technique has a number of advantages: (i) Two single electron states,
the \textbf{fair} states $a_{0}^{\dagger}$ and $b_{0}^{\dagger}$ determine the
full bases of the electron bands parallel and anti-parallel to the impurity
spin. Each \textbf{fair} state requires only a small number (of the order of
$40$) of coefficients $\alpha_{0}^{\nu}$ of Wilson states. (ii) The ground
states for the FA and the Kondo impurity consist of a small number of Slater
states. (iii) The d-state occupations in the different Slater states is well
separated insofar as each Slater state possesses either zero, one, or two d-electrons.

So far the error margins of the FAIR technique have not been quantified.
However, the quality of reproducing previous results justifies the use of this
rather transparent method to predict new phenomena and uncover relationships
and coherences which were not transparent before.

In this paper I denote the singlet ground state $\Psi_{SS}$ in the basis of
$N$ Wilson states as $\Psi_{N}$. For the FA-impurity $\Psi_{N}$ can be written
as a sum of (eight) Slater states with $N/2$ spin-up and $N/2$ spin-down
states. Each Slater state $S_{N}^{\dagger}$ is the product of $N/2$ spin-up
and $N/2$ spin-down electron states (creation operators applied to the vacuum
state $\Omega)$ and has a coefficient $\alpha_{S_{N}}$. I denote the sum of
all these Slater states including their coefficients as $A_{N}^{\dagger}%
\Omega=\Psi_{N}$. Similarly one can express the ground state for the $\left(
N-2\right)  $-Wilson basis as $\Psi_{N-2}=B_{N-2}^{\dagger}\Omega$. Then one
can multiply the state $\Psi_{N}$ with the adjoint operator of $B_{N-2}%
^{\dagger},$i.e the corresponding annihilation operators and form the state%
\begin{equation}
\left[  B_{N-2}^{\dagger}\right]  ^{\dagger}A_{N}^{\dagger}\Omega=B_{N-2}%
A_{N}^{\dagger}\Omega=\alpha_{Q}Q_{2}^{\dagger}\Omega\label{S_P}%
\end{equation}
This procedure yields a two-electron state $\alpha_{Q}Q_{2}^{\dagger}\Omega$
with one electron in each spin. ($\alpha_{Q}$ is the amplitude and
$Q_{2}^{\dagger}$ is normalized). In the final step I try to express the
two-electron state $Q_{2}^{\dagger}$ as the product of a spin-up and a
spin-down single-electron state $q_{\uparrow}^{\dagger}$ and $q_{\downarrow
}^{\dagger}$ with identical orbits,
\begin{equation}
\alpha_{Q}Q_{2}^{\dagger}\Omega=\alpha_{q}^{2}q_{\uparrow}^{\dagger
}q_{\downarrow}^{\dagger}\Omega\label{q_st}%
\end{equation}
The value $\alpha_{q}$ is the amplitude of the (normalized) state $q_{\sigma
}^{\dagger}$. The absolute value $\left\vert \alpha_{q}\right\vert $ is less
than or equal to one. This can be seen in the following way: We form the
$N$-particle state $B_{N-2}^{\dagger}Q_{2}^{\dagger}\Omega$ and take the
scalar product between $B_{N-2}^{\dagger}Q_{2}^{\dagger}\Omega$ and
$A_{N}^{\dagger}\Omega=\Psi_{N}$.
\[
\left\langle B_{N-2}^{\dagger}Q_{2}^{\dagger}\Omega|A_{N}^{\dagger}%
\Omega\right\rangle =\left\langle Q_{2}^{\dagger}\Omega|B_{N-2}A_{N}^{\dagger
}\Omega\right\rangle =\left\langle Q_{2}^{\dagger}\Omega|\alpha_{Q}%
Q_{2}^{\dagger}\Omega\right\rangle =\alpha_{Q}%
\]
where equ.(\ref{S_P}) is used. The scalar product is equal to $\alpha_{Q}$.
The absolute value of the scalar product between two normalized $N$-electron
states is less than or equal to one. Therefore one has $\left\vert \alpha
_{Q}\right\vert \leq1$. Under optimal conditions the two-electron state
$Q_{2}^{\dagger}$ can be factored into two single electron states, i.e.
$\alpha_{Q}Q_{2}^{\dagger}\simeq\left(  \alpha_{q}q_{\uparrow}^{\dagger
}\right)  \left(  \alpha_{q}q_{\downarrow}^{\dagger}\right)  \Omega.$ One
expects that the coefficient $\left\vert \alpha_{q}\right\vert $ is also less
than or equal to one.

After the two ground states $\Psi_{N}$ and $\Psi_{N-2}$ are constructed in
form of eight Slater states the calculation of $B_{N-2}A_{N}^{\dagger}\Omega$
consists mainly of scalar products between different $\left(  n-1\right)
$-electron states ($n=N/2$). The latter are determinants of single-electron
scalar products. The resulting coefficients of $\alpha_{Q}Q_{2}^{\dagger
}\Omega$ form a quadratic $\left(  N+1\right)  \times\left(  N+1\right)
$-matrix in terms of $N$ Wilson states plus one d-state for each spin. In the
last step the two-electron state $\alpha_{Q}Q_{2}^{\dagger}\Omega$ is split
into the product $\left(  \alpha_{q}q_{\uparrow}^{\dagger}\right)  \left(
\alpha_{q}q_{\downarrow}^{\dagger}\right)  \Omega$. This procedure is
remarkably easy. Already the square root of the diagonal elements in
$\alpha_{Q}Q_{2}^{\dagger}\Omega$ yields the absolute value of the
coefficients of $\left(  \alpha_{q}q_{\downarrow}^{\dagger}\right)  \Omega,$
and the sign follows from the non-diagonal elements.

Table I shows the value $\left\vert \alpha_{q}\right\vert $ for a number of
systems. The first column gives the impurity, the second the s-d-hopping
strength, and the third column the number $N$ of Wilson states. There are
three different values of $N$ used, $N=42,32$ and 22$.$The corresponding
values of the smallest energy spacing are $2^{-20}\approx10^{-6}%
,2^{-15}\approx3.\times\,\allowbreak10^{-5}$ and $2^{-10}\approx10^{-3}$.

For the FA impurities I used $U=1$ and $E_{d}=-0.5$. In all these cases the
split of the two-particle state $\alpha_{Q}Q_{2}^{\dagger}\Omega$ into two
identical single particle states with opposite spin worked very well. The
coefficient $\alpha_{q}$ in column four is very close to one. The fifth column
gives the Gaussian deviation $E_{rr}$ between $\alpha_{Q}Q_{2}^{\dagger}%
\Omega$ and $\left(  \alpha_{q}q_{\uparrow}^{\dagger}\right)  \left(
\alpha_{q}q_{\downarrow}^{\dagger}\right)  \Omega$. Obviously as soon as the
smallest energy in the Wilson spectrum $\delta E$ is sufficiently small
compared to the resonance width of the impurity at the Fermi level then the
state $q^{\dagger}$ is well defined and almost perfectly normalized. This
means that the state $\Psi_{N}$ can be constructed from the state $\Psi_{N-2}$
by the relation%
\begin{equation}
\Psi_{N}\cong q_{\uparrow}^{\dagger}q_{\downarrow}^{\dagger}\Psi_{N-2}
\label{recurs}%
\end{equation}

\begin{align*}
&
\begin{tabular}
[c]{|l|l|l|l|l|l|l|}\hline
\textbf{impurity} & $\left\vert \mathbf{V}_{sd}\right\vert ^{2}$ &
$\mathbf{N}$ & $\left\vert \mathbf{\alpha}_{q}\right\vert $ & $E_{rr}$ &
$c_{q}\left(  -1\right)  $ & c$_{q}\left(  0\right)  $\\\hline
Friedel & $10^{-4}$ & $42$ & $0.995$ & $10^{-13}$ & $0.814$ & $-0.528$\\\hline
Friedel & $0.03$ & $22$ & $0.995$ & $10^{-13}$ & $0.813$ & $-0.529$\\\hline
FA & $0.03$ & $42$ & $0.990$ & $3\times10^{4}$ & $0.814$ & $-0.525$\\\hline
FA & $0.04$ & $42$ & $0.998$ & $5\times10^{-5}$ & $0.814$ & $-0.528$\\\hline
FA & $0.05$ & $22$ & $0.979$ & $2\times10^{3}$ & $0.812$ & $-0.504$\\\hline
FA & $0.05$ & $32$ & $0.997$ & $2\times10^{-4}$ & $0.814$ & $-0.527$\\\hline
FA & $0.05$ & $42$ & $0.998$ & $2\times10^{-5}$ & $0.812$ & $-0.533$\\\hline
Kondo & $J=0.10$ & $42$ & $0.992$ & $2\times10^{-4}$ & $0.8062$ &
$-0.533$\\\hline
\end{tabular}
\\
&
\begin{tabular}
[c]{l}%
Table I: The first column gives the impurity, the second the\\
s-d-hopping strength, the third column the number $N$ of Wilson\\
states. The fourth column gives the coefficient $\left\vert \alpha
_{q}\right\vert $ of the\\
extracted state $q^{\dagger}$. The next column gives the Gaussian\\
deviation $E_{rr}.$The following two columns are explained in the text.
\end{tabular}
\end{align*}%
\[
\]

In all cases the extracted state $q^{\dagger}$ is mainly composed of basis
states close the Fermi level. In table I the two last columns give the
amplitudes of the state $q^{\dagger}$ below and at the new Fermi level in
terms of Wilson states. These amplitudes agree almost perfectly.

Fig.3 shows the coefficients of $q^{\dagger}$ with respect to the Fermi level
for pure Friedel impurities with $N=42$ and very different s-d-hopping
strengths $\left\vert V_{sd}\right\vert ^{2}$. It also shows the corresponding
coefficients of $q^{\dagger}$ for FA impurities with $N=42$ and different
$\left\vert V_{sd}\right\vert ^{2}$ (which yield many orders of magnitude
different Kondo energies). In addition Fig.3 shows the effect of different $N$
for a FA impurity with $\left\vert V_{sd}\right\vert ^{2}=0.05$, $U=1$ and
$E_{d}=-0.5$. The Kondo energy is about $3.9\times10^{-3}$. For $N=22$ the
smallest energy spacing is about $10^{-3}$. Still the $N=22$ curve is
relatively close to the universal curve although table I shows that its
$\left\vert \alpha_{q}\right\vert =0.979$ is less close to one than for larger
$N$. For values of $N$ where the smallest level spacing is of the order of the
Kondo energy or larger the value of $\alpha_{q}$ decreases. Then the state
$\Psi_{N}$ can no longer be expressed as $q_{\uparrow}^{\dagger}q_{\downarrow
}^{\dagger}\Psi_{N-2}$ and the splitting of $Q_{2}^{\dagger}$ into two single
electron states with opposite spin becomes meaningless.
\begin{align*}
&
\raisebox{-0pt}{\includegraphics[
height=3.2312in,
width=4.005in
]%
{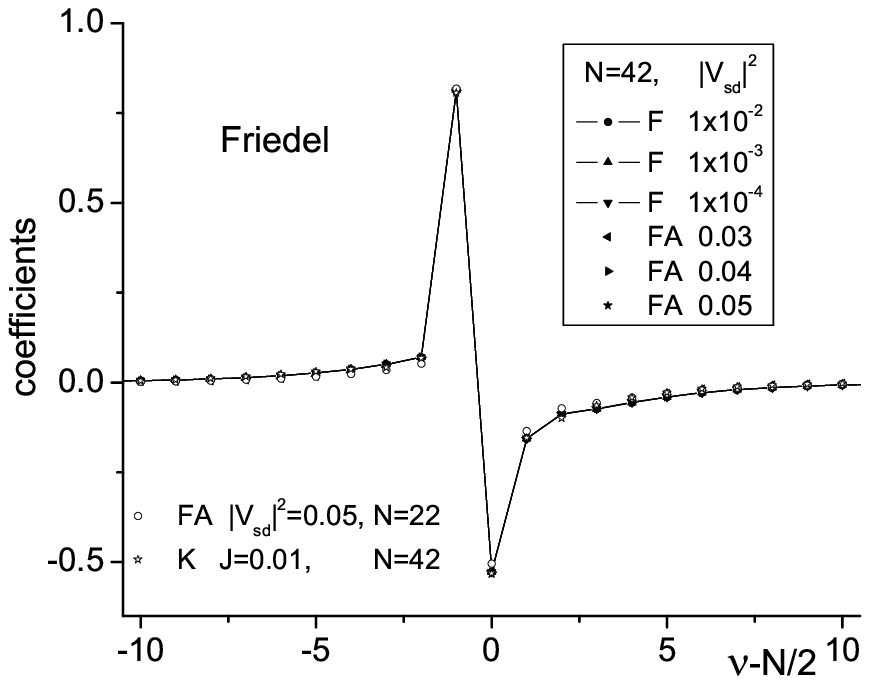}%
}%
\\
&
\begin{tabular}
[c]{l}%
Fig.3: This figure shows the coefficients of the state $q^{\dagger}$ for\\
three different impurities, F =Friedel, FA=Friedel-Anderson\\
and K=Kondo. The coefficients are counted from the Fermi\\
level. Most curves are for $N=42,$ but one curve is for a\\
FA impurity with $\left\vert V_{sd}\right\vert ^{2}=0.05$ and $N=22$. The
curve shown\\
is super-universal because it is independent of the kind of\\
impurity (F,FA,K), independent of $N,$ and independent of $\left\vert
V_{sd}\right\vert ^{2}$\\
and other parameters as long as there is a resonance exactly\\
at the Fermi level.
\end{tabular}
\end{align*}%
\[
\]

I also performed the same calculation for a Kondo impurity. The parameters are
collected in table I as well. The same behavior is observed. Fig.3 includes
the coefficients of a Kondo impurity with $J=0.10$ and $N=42$ as open stars.

The interpretation of the above results is that the complexity of the
solutions of the FA and the Kondo impurity is not felt at energies
sufficiently below the Kondo energy. We know from Wilson's renormalization
calculation that the structure of the ground state changes dramatically when
the smallest level spacing $\delta E$ is of the order of the Kondo energy.
After this metamorphosis the renormalization approaches a fixed point. From
this behavior Nozieres \cite{N14} developed the Fermi liquid theory of the
Kondo impurity (which applies also to the FA impurity). The present
calculation provides quantitatively a universal state $q_{\sigma}^{\dagger}$
which has to be incorporated into the ground state for both spins when the
Wilson basis is increased by 2 states. The fact that this is the same state
for any impurity with a resonance at the Fermi level demonstrates that the
Kondo and FA impurities have a resonance at the Fermi level. The result
suggests an induction rule for the ground state with increasing number of
Wilson states.

The accuracy of relation (\ref{recurs}) is given by the scalar product between
the left and right side. This is for the FA impurity and $N=42$ already 0.996
$($i.e equal to $\left\vert \alpha_{q}\right\vert ^{2})$ and approaches the
value one with increasing $N$.

The fact that $q^{\dagger}$ is the same state for any impurity with a
resonance at the Fermi level is at the heart of our earlier discussed fidelity
calculations. The fidelity is the scalar product between the ground states of
the symmetric Friedel and FA impurities, $\left\langle \Psi_{N}^{F}|\Psi
_{N}^{FA}\right\rangle $, which is compared with the corresponding scalar
product in the $\left(  N-2\right)  $-basis $\left\langle \Psi_{N-2}^{F}%
|\Psi_{N-2}^{FA}\right\rangle $. (In both cases the bands are half-filled).
According to our present result we can express the latter as
\[
\left\langle \Psi_{N}^{F}|\Psi_{N}^{FA}\right\rangle \cong\left\langle
q_{\uparrow}^{\dagger}q_{\downarrow}^{\dagger}\Psi_{N-2}^{F}|q_{\uparrow
}^{\dagger}q_{\downarrow}^{\dagger}\Psi_{N-2}^{FA}\right\rangle =\left\langle
\Psi_{N-2}^{F}|\Psi_{N-2}^{FA}\right\rangle
\]
This relation confirms that the fidelity between the symmetric FA and the
symmetric Friedel impurity approaches a finite value with increasing $N$ and
does not experience an Anderson orthogonality catastrophe. The equation
(\ref{recurs}) yields an asymptotic recursion formula to construct $\Psi_{N}$
from a given $\Psi_{N_{0}}$where $N_{0}$ is sufficiently large so that the
smallest level spacing in the $N_{0}$-Wilson basis is smaller than the Kondo
energy.%
\[
\]

Conclusion: In the Wilson basis one can easily introduce two new states by
splitting each of the states directly below and above the center of the band
into two states of equal weight, creating out of two old states four new ones.
This reduces the smallest energy spacing by a factor of two and increases the
effective number of states $N_{eff}$ by a \textbf{factor} of two. The relation
between between the ground states $\Psi_{N-2}$ and $\Psi_{N}$ of the two half
filled bands is investigated for several symmetric impurities, the Friedel,
the Friedel-Anderson and the Kondo impurity. The ground states are calculated
with the FAIR technique. We observe that for sufficiently large $N$ the ground
state $\Psi_{N}$ can be obtained from $\Psi_{N-2}$ by multiplying $\Psi_{N-2}$
by two single particle states $q_{\uparrow}^{\dagger}$ and $q_{\downarrow
}^{\dagger}$. The composition of these states $q^{\dagger}$with respect to the
Fermi level is independent of $N$ (for sufficient large $N$) and is
independent of the resonance width of the impurity. This qualifies the state
$q^{\dagger}$ to be called universal. The state $q^{\dagger}$ is not just
universal for a given kind of impurity. It is super-universal because it is
even independent of the kind of impurity as long as the impurity has a
resonance directly at the Fermi level. In particular it is independent of the
complexity of the ground state of the FA or the Kondo impurity. All the
complexity happens further away from the Fermi level. This result gives a
simple confirmation and explanation of our previous fidelity calculations
\cite{B185}. A similar investigation for the asymmetric FA impurity is in
progress. %

\[
\]
Acknowledgement: The author acknowledges the stimulating comment of one of the
referees of ref. \cite{B185}.


\newpage

\begin{thebibliography}{99}                                                                                               %


\bibitem {A53}P. W. Anderson, Phys. Rev. Lett. 18, 1049 (1967)

\bibitem {F28}J. Friedel, Adv. Phys. 3, 446 (1954); Can. J. Phys. 34, 1190
(1956); Suppl. Nuovo Cimento 7, 287 (1958); J. Phys. Radium 19, 573 (1958)

\bibitem {A31}P. W. Anderson, Phys. Rev. 124, 41 (1961)

\bibitem {S88}W. Brenig and K. Schoenhammer, Z. Physik 267, 201 (1974)

\bibitem {L57}D. E. Logan, M. P. Eastwood, and M. A. Tusch, J. Phys.: Condens.
Matter 10, 2673 (1998)

\bibitem {B181}G. Bergmann, Eur. Phys. J. B 75, 497 (2010)

\bibitem {H20}A. C. Hewson, The Kondo problem to heavy Fermions, Cambridge
University Press, 1993

\bibitem {W50}A. Weichselbaum, W. M\"{u}nder, J. v. Delft, Phys. Rev. B 84,
075137 (2011)

\bibitem {B185}G. Bergmann and R. S. Thompson, Eur. Phys. J. B 84, 273 (2011)

\bibitem {W12}P. B. Wiegmann, in Quantum Theory of Solids, edited by I. M.
Lifshitz (MIR Publishers, Moscow, 1982), p. 238

\bibitem {S29}P. Schlottmann, Phys. Reports 181, 1 (1989)

\bibitem {W18}K. G. Wilson, Rev. Mod. Phys. 47, 773 (1975)

\bibitem {K58}H. R. Krishna-murthy, J. W. Wilkins, and K. G. Wilson, Phys.
Rev. B 21, 1003 (1980)

\bibitem {B91}G. Bergmann, Z. Physik B102, 381 (1997)

\bibitem {B151}G. Bergmann, Phys. Rev. B 74, 144420 (2006)

\bibitem {B153}G. Bergmann and L. Zhang, Phys. Rev. B 76, 064401 (2007)

\bibitem {B177}G. Bergmann, Phys. Rev. B 77, 104401 (2008)

\bibitem {B182}G. Bergmann, and Y. Tao, Eur. Phys. J. B 73, 95 (2010)

\bibitem {B186}Y. Tao and G. Bergmann, Eur. Phys. J. B 85, 42 (2012)

\bibitem {B152}G. Bergmann, Phys. Rev. B 73, 092418 (2006)

\bibitem {B187}G. Bergmann, J. Supercond. Nov. Magn. 25, 609 (2012)

\bibitem {N14}P. Nozieres, J. Low Temp. Phys. 17, 31 (1974)
\end{thebibliography}
\end{document}